\documentclass[twocolumn,amsmath,aps,fleqn]{revtex4}
\usepackage{graphicx,amssymb}
\begin{document}
\newcommand{\be}{\begin{equation}}
\newcommand{\ee}{\end{equation}}
\newcommand{\bq}{\begin{eqnarray}}
\newcommand{\eq}{\end{eqnarray}}
\newcommand{\bsq}{\begin{subequations}}
\newcommand{\esq}{\end{subequations}}
\newcommand{\bc}{\begin{center}}
\newcommand{\ec}{\end{center}}
\newcommand {\R}{{\mathcal R}}
\newcommand{\al}{\alpha}
\newcommand\lsim{\mathrel{\rlap{\lower4pt\hbox{\hskip1pt$\sim$}}
    \raise1pt\hbox{$<$}}}
\newcommand\gsim{\mathrel{\rlap{\lower4pt\hbox{\hskip1pt$\sim$}}
    \raise1pt\hbox{$>$}}}

\title{Constraints on the dark matter sound speed from galactic scales: the cases of the Modified and Extended Chaplygin Gas}

\author{P. P. Avelino}
\email[Electronic address: ]{pedro.avelino@astro.up.pt}
\affiliation{Instituto de Astrof\'{\i}sica e Ci\^encias do Espa{\c c}o, Universidade do Porto, CAUP, Rua das Estrelas, PT4150-762 Porto, Portugal}
\affiliation{Centro de Astrof\'{\i}sica da Universidade do Porto, Rua das Estrelas, PT4150-762 Porto, Portugal}
\affiliation{Departamento de F\'{\i}sica e Astronomia, Faculdade de Ci\^encias, Universidade do Porto, Rua do Campo Alegre 687, PT4169-007 Porto, Portugal}

\author{V. M. C. Ferreira}
\email[Electronic address: ]{vascoferreira1@gmail.com}
\affiliation{Instituto de Astrof\'{\i}sica e Ci\^encias do Espa{\c c}o, Universidade do Porto, CAUP, Rua das Estrelas, PT4150-762 Porto, Portugal}
\affiliation{Centro de Astrof\'{\i}sica da Universidade do Porto, Rua das Estrelas, PT4150-762 Porto, Portugal}

\date{\today}
\begin{abstract}

We show that the observed rotation curves of spiral galaxies constrain the sound speed of the dark matter to be $c_s < 10^{-4} c$, where $c$ is the speed of light in vacuum. Using the Modified Chaplygin Gas as a representative example of a class of unified dark energy models incorporating an effective dark matter component with a non-zero sound speed, we determine the most stringent constraint to date on the value of the constant contribution to the equation of state parameter in this class of models. Finally, we explain the reason why previous constraints using the Cosmic Microwave Background and Baryonic Acoustic Oscillations were not as competitive as the one presented in this paper and discuss the limitations of the recently proposed Extended Chaplygin Gas.

\end{abstract}
\maketitle

\section{\label{intr}Introduction}

Increasingly precise cosmological observations \cite{Suzuki:2011hu,Anderson:2012sa,Parkinson:2012vd,Hinshaw:2012aka,Ade:2013zuv,Planck:2015xua} remain consistent with the standard $\Lambda$CDM model. In this model it is assumed that general relativity provides an accurate description of gravity on cosmological scales, and that the two main constituents of the Universe are a cosmological constant $\Lambda$ and a non-relativistic Dark Matter (DM) component. While a cosmological constant, or a more general Dark Energy (DE) form, is necessary in the context of general relativity to account for the current acceleration of the Universe, the DM component is required in order to explain the observed dynamics of cosmological perturbations over a wide range of scales. Despite the simplicity of the cosmological constant, there is presently no satisfactory explanation for its tiny energy density, which favours dynamical DE or modified gravity as a better motivated explanation for the acceleration of the Universe (see, for example \cite{Copeland:2006wr,Frieman:2008sn,Caldwell:2009ix,Li:2011sd,Bamba:2012cp} and references therein). In fact there is no alternative to dynamical DE or modified gravity if the vacuum energy is screened, and prevented from acting as a gravitational source (see, e.g., \cite{Kaloper:2013zca,Avelino:2014nqa}).

Another (albeit equivalent) interpretation of the $\Lambda$CDM model relies on a single perfect fluid with constant negative pressure. Hence, $\Lambda$CDM may be regarded as the simplest Unified Dark Energy (UDE) model, in which DM and DE are taken as different manifestations of a single unified DE fluid \cite{Avelino:2003cf}. A broad class of UDE models, known as the Generalized Chaplygin Gas (GCG) \cite{Bento:2002ps}, includes the original Chaplygin Gas \cite{Bilic:2001cg,Gorini:2002kf} and $\Lambda$CDM as particular models. The GCG has been claimed to be essentially ruled out due to the late time oscillations or the exponential blow up of the DM power spectrum predicted using linear perturbation theory \cite{Sandvik:2002jz} (except for a tiny region of parameter space very close to the $\Lambda$CDM limit). However, it has recently been show that the GCG may be consistent with current observational constraints, over a wide region of parameter space, assuming that there is a sufficiently high level of non-linear clustering on small scales \cite{Avelino:2014nva} (see also \cite{Avelino:2003ig,Beca:2005gc,Avelino:2007tu,Avelino:2008cu}).

The Modified Chaplygin Gas (MCG) has been proposed as a further generalization of the GCG model \cite{Benaoum:2002zs}. This model incorporates an extra parameter which is associated to a lower bound on the sound speed of the effective DM component. The MCG  has been observationally constrained using the redshift dependence of the apparent magnitude of type Ia supernovae, the Cosmic Microwave Background (CMB) anisotropies, and the matter power spectrum, including Baryonic Acoustic Oscillations (BAO) \cite{Liu:2005dk,Balbi:2007mz,Lu:2008zzb,Thakur:2009jg,Fabris:2010vd,Lu:2010zzj,Paul:2012zz,Xu:2012ca,Paul:2013sha,Panigrahi:2013ara}. Further extensions incorporating additional terms accounting for a possible growth of the comoving Jeans length with cosmic density have also been proposed in the literature (see, e.g., \cite{Pourhassan:2014ika}).

In this paper we shall constrain the value of the DM sound speed using the observed rotation curves of spiral galaxies, and use that result to impose stringent limits on the MCG class of models.  A comparison with previous constraints obtained using other observational data and a discussion of the limitations of the recently proposed Extended Chaplygin Gas (ECG) will also be presented. 

Throughout this paper we use units in which the speed of light in vacuum is $c=1$.

\section{Modified Chaplygin Gas\label{sec2}}

The MCG can be described as a perfect fluid with an energy-momentum tensor
\be
T^{\mu\nu}=(\rho+p)u^\mu u^\nu + p g^{\mu \nu}\,,
\ee
having a barotropic equation of state parameter given by
\be
w=B-\frac{A}{\rho^{\alpha+1}}\,,
\ee
where $A$, $B$ and $\alpha$ are constant parameters, $w=p/\rho$, $\rho$ is the proper energy density, $p$ is the hydrostatic pressure, $u^\mu$ are the components of the 4-velocity and $g^{\mu \nu}$ are the components of the metric tensor.

For $B=0$ the MCG reduces to the GCG, which is arguably the simplest non-trivial class of models where the role of DM and DE is played by a single dark fluid. On the other hand, for $B=0$ and $\alpha=0$ the MCG is completely equivalent to the $\Lambda$CDM model. In this paper we shall mainly be interested in the case where $B \neq 0$, in which case the sound speed is given by 
\be
c_s^2=\frac{dp}{d\rho}=B+\alpha\frac{A}{\rho^{\alpha+1}}\,.
\ee
Since imaginary sound speeds are associated with instabilities, in particular on very small scales, in this paper we shall consider model parameters in the range $A \ge 0$, $B \ge  0$ and $0 \le \alpha \le 1$, for which $c_s^2 \ge 0$. Note that the value of $c_s^2$ is always greater than or equal to $B$ but it tends to $B$ for large enough energy densities. In this limit the MCG behaves effectively as DM.

\section{Observational constraints\label{sec3}}

\subsection{Galactic rotation curves}

The observed rotation curves of spiral galaxies provide striking evidence for the presence of DM on galactic scales (see, e.g., \cite{Sofue:1999jy,Sofue:2000jx,Walter:2008wy}). These curves typically flatten at large distances from the galactic center, around and well beyond the edge of the visible disks, indicating the presence of a (nearly) spherically symmetric DM halo. In the Newtonian limit, expected to be valid on galactic scales, the circular velocity $v$ is given by
\be
v^2(r)=\frac{GM(r)}{r}=\frac{4 \pi}{3} G \rho(r) r^2\,,
\ee
so that
\be
G \rho(r)= \frac{3}{4 \pi} \frac{v_f^2}{r^2}\,,\label{Grho}
\ee
in the flat part of the spiral galaxy rotation curves. Here $r$ is the distance from the galactic center, $G$ is the gravitational constant, $M(r)$ is the total mass inside a sphere of radius $r$ centered in $r=0$, $\rho(r)$ is the average density inside that sphere, and $v_f$ is the value of the circular velocity at large distances from the galactic center.

If the DM has a non-zero sound speed, then its gravitational collapse can only take place on scales larger than the Jeans length \cite{Jeans:1902}
\be
\lambda_{J[DM]}=c_{s[DM]}{\sqrt{\frac{\pi}{G\rho}}}\,.\label{lambdaJDM}
\ee
On smaller scales the DM pressure balances gravity giving rise to stable oscillations.

Substituting Eq. (\ref{Grho}) into Eq. (\ref{lambdaJDM}) one obtains
\be
\frac{\lambda_{J[DM]}}{r}= {\sqrt {\frac{4}{3}}} \pi \frac{c_{s[DM]}}{v_f}\,.
\ee
DM may only collapse on length scales smaller than $r$ if $\lambda_J < r$ or, equivalently, if
\be
c_{s[DM]} < {\sqrt {\frac{3}{4}}} \frac{v_f}{\pi} \,.
\ee
In the context of the MCG UDE model, this leads to the following constraint 
\be
B < c_{s[DM]}^2 < \frac34 \frac{v_f^2}{\pi^2}\,.
\ee
Here, $c_{s[DM]}$ is to be interpreted as the sound speed of the MCG in the galactic halo (again note that the MCG sound speed squared approaches $B$ at high energy densities). Taking into account that the typical values of $v_f$ observed for spiral galaxies are in the range $100\, {\rm km \, s^{-1}}\lsim v_f \lsim 300\, {\rm km \, s^{-1}}$ \cite{Sofue:1999jy,Sofue:2000jx,Walter:2008wy}, with some galaxies having lower maximum velocities, one finally obtains the following conservative limit on the value of $B$
\be
B < 10^{-8}\,. \label{Bmax}
\ee

\subsection{Other observational constraints}

The constraint given in Eq. (\ref{Bmax}) is much more stringent than those obtained using the CMB or BAO (see, e.g., \cite{Xu:2012ca}). Here we shall explain why. 

The adiabatic sound speed of the baryon-photon ($b\gamma$) plasma before last scattering is given by \cite{Hu:1996vq}
\be
c_{s[b\gamma]}^2=\frac{1}{3}\left(1+\frac{3}{4}\frac{\rho_b}{\rho_\gamma}\right)^{-1}=\frac13\left(1+\frac{3}{4}\frac{\Omega_{b0}}{\Omega_{\gamma 0}(1+z)}\right)^{-1}\,,
\ee
where $z=1-1/a$ is the redshift, $a$ is the cosmological scale factor, $\Omega_i=\rho_i/\rho_c$, an `$i$' represents a particular energy component, in this case either baryons ($b$) or photons ($\gamma$), $\rho_c=3 H^2/8\pi G$ is the critical density, $H= 100 \, h \, {\rm km} \, {\rm s}^{-1} \, {\rm Mpc}^{-1}$ is the Hubble parameter and a `0' denotes the present time. Using the constraints on the values of the cosmological parameters obtained by the Planck 2015 team, namely $\Omega_{b0} h^2 = 0.02230 \pm 0.00014$, $z_{rec}=1089.90 \pm 0.23$ and $h=0.6774 \pm 0.0046$ \cite{Planck:2015xua}, as well as $\Omega_{\gamma 0} h^2=2.47 \times 10^{-5}$ one may estimate the sound speed of the baryon photon plasma just before recombination
\be
c_{s[b\gamma]}^2(z_{rec-}) =\frac{1}{3}\left(1+\frac{3}{4}\frac{\Omega_{b0}}{\Omega_{\gamma0}\left(1+z_{rec}\right)}\right)^{-1} \sim 2 \times 10^{-1}\,.
\ee
Here $z_{rec}$ denotes the redshift of recombination and a minus represents an instant immediately before that. The position of the first acoustic peak at $\ell \sim 200$ is extremely sensitive to the value of $c_{s[b\gamma]}(z_{rec-})$.

A non-zero DM sound speed would be responsible for additional signatures in the CMB primary anisotropies. In order for them to only affect multipoles with $\ell >  2500$ ($\ell \sim 2500$ is roughly the upper limit of the multipole range probed by Planck, and twice that of WMAP), the value of the DM sound speed would need to be 
\bq
c_{s[DM]}^2(z_{rec-})  <  \left(\frac{200}{2500}\right)^2  c_{s[b\gamma]}^2(z_{rec-}) \sim 1 \times 10^{-3}\,,
\eq
or, equivalently, $B<10^{-3}$ if the GCG is considered as a UDE model. Not too surprisingly, this limit is of the same order as that otained in \cite{Xu:2012ca} using the WMAP seven year CMB data (together with BAO and type Ia Supernovae data). Note that the Integrated Sachs-Wolfe effect has also a significant contribution to the CMB constraints reported in \cite{Xu:2012ca} (see also \cite{Bertacca:2007}).

Given that the Planck results constrain the matter power spectrum on comoving wavenumbers $k \lsim 0.2 \, {\rm Mpc}^{-1}$ \cite{Ade:2015oja}, the current CMB data is only able to constrain the DM sound speed at $z_{rec-}$ on comoving scales larger than $\lambda \gsim 30 \, {\rm Mpc}$. Since the evolution of the DM cosmological comoving Jeans length during the matter dominated era is given by
\be
\lambda_{J[DM]}^c \equiv \frac{\lambda_{J[DM]}}{a} \propto \frac{c_{s[DM]}}{\rho^{1/2}a} \propto c_{s[DM]} (1+z)^{-1/2}\,,
\ee
it is possible to improve the CMB constraints on the value of a constant $c_{s[DM]}^2$ (or $B$) by a factor of approximately $1+z_{rec} \sim 10^3$ using the observed matter power spectrum  at $z=0$ for comoving wavenumbers $k \lsim 0.2 \, {\rm Mpc}^{-1}$. This is consistent with the constraint on the GCG class of models reported in \cite{Fabris:2010vd} ($B<10^{-6}$). On the other hand, Lyman-alpha constraints on the matter power spectrum on comoving wavenumbers up to $k \sim 2 \, {\rm Mpc}^{-1}$ \cite{McDonald:2004eu,Palanque-Delabrouille:2013gaa} are expected to lead to even tighter limits on the value of $B$, only slightly less constraining than the ones obtained in the present paper using the galactic rotation curves.

High redshift constraints on the value of the DM sound speed may be more effective than low redshift ones in models where the Jeans length grows with redshift. An example of a class of models where this can be realized is the ECG, in which the equation of state parameter is given by \cite{Pourhassan:2014ika}
\be
w=\sum_n B_n \rho^{n-1} -\frac{A}{\rho^{\alpha+1}}\,.
\ee
Here, $n>0$ are integers and $B_n$ are real constants. For the sake of simplicity let us assume that $A=0$, $B_2 > 0$ and $B_n=0$ for $n \neq 2$. In this case, the ECG sound speed is given by
\be
c_s^2=2  B_2 \rho \propto (1+z)^3\,,
\ee
implying that the cosmological comoving Jeans length is 
\be
\lambda_{J[DM]}^c \equiv \frac{\lambda_{J[DM]}}{a} \propto \frac{c_{s[DM]}}{\rho^{1/2}a}  \propto \frac{\rho^{1/2}}{a} \propto (1+z)^{5/2}\,,
\ee 
during the matter dominated era. This leads to a value of $\lambda_{J[DM]}^c(z_{rec-})$ which is larger by a factor of more than $10^7$ than the value at $z=0$. As a consequence, for $n \ge 2$ CMB constraints on the ECG class of models effectively rule nearly all the available parameter space, except for a very small region with $B_n \sim 0$.

\section{\label{conc} Conclusions}

In this paper we have determined the most stringent constraint to date on the effective DM sound speed using the  observed circular velocity of spiral galaxies at large distances from the galactic centre. Our results were then used to show that $B < 10^{-8}$ if the MCG is to be regarded as a UDE  candidate. We have compared this constraint with those obtained by various authors using other observational data, explaining the reason for the significant improvement obtained in this paper. Finally, we have discussed the case of the ECG, as an example of a class of models where the Jeans length may increase significantly with redshift, showing that, in this case, CMB constraints are extremely effective at constraining the available parameter space.

\begin{acknowledgments}

P.P.A. is supported by Funda{\c c}\~ao para a Ci\^encia e a Tecnologia (FCT) through the Investigador FCT contract of reference IF/00863/2012 and POPH/FSE (EC) by FEDER funding through the program "Programa Operacional de Factores de Competitividade - COMPETE.
\end{acknowledgments}


\bibliography{MCG}

\end{document}